\documentclass{article}[a4]

\usepackage{amsmath}
\usepackage{amssymb}
\usepackage{authblk}
\usepackage[english]{babel}
\usepackage[
    sortcites,
    backend=biber,
    giveninits=true,
    maxbibnames=5,
    ]{biblatex}
\usepackage[utf8]{inputenc}
\usepackage{csquotes}
\usepackage[T1]{fontenc}
\usepackage{microtype}
\usepackage{pgfplots}
\usepackage{subfig}
\usepackage{tikz}

\usetikzlibrary{arrows.meta}
\usetikzlibrary{backgrounds}
\usetikzlibrary{fit}
\usetikzlibrary{shapes}

\pgfplotsset{compat=1.5}

\pgfplotstableread{
length count
0 3
1 2
2 1
3 1
4 1
5 2
6 2
7 2
8 4
9 3
10 3
11 4
12 4
13 6
14 6
15 9
16 10
17 9
18 16
19 7
20 10
21 10
22 19
23 16
24 18
25 16
26 11
27 16
28 18
29 33
30 59
31 30
32 43
33 14
34 14
35 31
36 19
37 15
38 18
39 10
40 24
41 21
42 26
43 56
44 36
45 20
46 68
47 64
48 90
49 67
50 37
51 41
52 20
53 4
54 13
55 30
56 39
57 36
58 16
59 11
61 13
62 131
63 22
66 17
67 66
68 26
72 109
73 168
74 27
75 23
77 18
78 31
79 43
80 16
83 18
90 36
91 37
94 34
113 25
137 57
150 84
151 77
160 113
187 96
238 98
256 147
270 181
367 250
412 237
534 332
1333 365
1462 435
}\influencelength

\pgfplotstableread{
relations count
1 16735
2 5589
3 2081
4 978
5 500
6 366
7 254
8 223
9 126
10 138
11 1472
}\relationscount

\newcommand*\circled[1]{\raisebox{.5pt}{\textcircled{\raisebox{-.9pt} {#1}}}}
\newcommand\scalemath[2]{\scalebox{#1}{\mbox{\ensuremath{\displaystyle #2}}}}

\newtheorem{definition}{Definition}[section]

\newcommand{\tikzmark}[1]{\tikz[overlay, remember picture] \node (#1) {};}

\DeclareRobustCommand{\changeStar}{
\begin{tikzpicture}[yscale=-.8, baseline=-1mm]%
\node[rectangle, fill=blue!30, inner sep=0pt, minimum height=4mm, minimum width=4mm] {$\star$};%
\end{tikzpicture}
}%

\DeclareRobustCommand{\changeDagger}{
\begin{tikzpicture}[yscale=-.8, baseline=-1mm]%
\node[rectangle, fill=orange!30, inner sep=0pt, minimum size=4mm, minimum width=4mm] {$\dagger$};%
\end{tikzpicture}
}%

\AtBeginBibliography{\small}
\author[1,2,*]{Jonathan K. Vis}
\author[1,3]{Mark A. Santcroos}
\author[2]{Walter A. Kosters}
\author[1,4]{Jeroen F.J. Laros}
\affil[1]{Department of Human Genetics, Leiden University Medical Center}
\affil[2]{Leiden Institute of Advanced Computer Science, Leiden University}
\affil[3]{Department of Clinical Genetics, Leiden University Medical Center}
\affil[4]{National Institute for Public Health and the Environment (RIVM)}

\affil[*]{e-mail: \texttt{j.k.vis@lumc.nl}}

\title{A Boolean Algebra for Genetic Variants}

\begin{document}
\maketitle

\begin{abstract}
\noindent
Beyond identifying genetic variants, we introduce a set of Boolean
relations that allows for a comprehensive classification of the relations
for every pair of variants by taking all minimal alignments into account.
We present an efficient algorithm to compute these relations, including a
novel way of efficiently computing all minimal alignments within the best
theoretical complexity bounds. We show that for variants of the
\textit{CFTR}~gene in dbSNP these relations are common and many
non-trivial. Ultimately, we present an approach for the storing and
indexing of variants in the context of a database that enables
efficient querying for all these relations.
\end{abstract}

\section{Introduction}\label{sec:intro}

DNA-sequencing aims to measure the genetic makeup of individuals.
Without going into details about the many different technologies, these
processes determine (fragments of) the genetic sequence. Commonly, the
primary data analysis consists, among other steps, of: 1) alignment
against a reference genome, e.g., GRCh38 for human samples, and 2)
variant calling. The primary result is a list of variants, i.e., a set of
differences that is specific for the measured individual (sample), often
reported in a tabular file like the Variant Call Format~(VCF)~\cite{vcf}.
These variants are used in subsequent applications ranging from
fundamental and association research studies to clinical diagnostics. It
is advantageous to look only at differences (with regard to some
reference) as the genome is usually large~(ca.~$3 \cdot 10^9$ nucleotides
for humans), but the individual differences between two genomes are
relatively small~(ca.~$0.6\%$~\cite{variation}).

When variants are associated with phenotypic traits, they are reported in
literature and stored with their annotation in (locus specific)
databases. Usually, the representation of the variant in VCF is refined
to a representation more suitable for reporting.
For this, many (domain specific) languages exist. Most notable are:
\begin{itemize}
\item Recommendations of the Human Genome Variation
Society~(HGVS)~\cite{hgvs};
\item SPDI~\cite{spdi}, the internal data model for variants used by the
National Center for Biotechnology Information (NCBI);
\item Variant Representation Specification~(VRS)~\cite{vrs}, developed by
the Global Alliance for Genomic Health (GA4GH).
\end{itemize}
These languages attempt to represent the observed differences in a
human-understandable and/or machine interpretable manner and, whereas VCF
is implicitly tied to the tooling and configuration used in the primary
data analysis, these representations are process agnostic and universally
interpretable.

Within the domain of variant recording, some simplifications are common.
First, small (local) variants on a single molecular sequence (part of the
same haplotype) are recorded separately, because this is convenient when
storing large numbers of variants in databases. Phasing information,
i.e., whether small variants are part of the same haplotype, is often
lost or incomplete. This is partially a direct consequence of the
sequencing technology and partially because this information is removed.
Second, in some representations (notably, HGVS) uncertainties might be
expressed. Usually, the uncertainties relate to the positioning of the
variant within the reference genome, but also the exact makeup of larger
insertions might by unknown.
Finally, unchanged regions may be implicit. During primary data analysis,
in particular the alignment step, the sequence from the reference genome
is assumed to be present even when direct evidence, e.g., coverage
information from the sequencing process, is lacking.

For the remainder of this paper, we adopt a strict view on the nature of
variants:
\begin{enumerate}
\item A variant consists of deletions, insertions or a
combination thereof with respect to a single molecular sequence.
When these operations occur in combination, they are said to be phased,
in cis or part of the same allele and can be written down as phase sets
or allele descriptions.
Many variant description languages have introduced higher-order
operations like single nucleotide variants (SNV) (called substitutions in
HGVS), multi-nucleotide variants (deletion/insertions), duplications,
transpositions, inversions, repeats, etc. We consider all of these
notions to be special cases of the definition given above.
\item We consider only \emph{interpretable} variants, i.e., given a
\emph{reference sequence}, there is a deterministic and unambiguous way
of ``applying'' the variants such that the result is the (originally)
measured \emph{observed sequence}, cf. the Unix~\texttt{diff} and
\texttt{patch} utilities.
\end{enumerate}

\noindent
As is already observed within the various variant representation
languages, it is often possible to have multiple representations
describing the same observed sequence. These possibilities can originate
from the choice of ``operator'', e.g., an SNV can also be represented by
a deletion of one nucleotide followed by an insertion of again one
nucleotide. Another source contributing to the number of possibilities
is the structure of the reference sequence.
Consider the reference sequence~\texttt{ATTTA} and the observed
sequence~\texttt{ATTA}. One of the symbols~\texttt{T} is removed; to say
which one specifically yields a number~($3$) of possibilities.
To determine a universally accepted representation of a variant, most
variant representation languages employ a \emph{normalization} procedure.
Normalization chooses a \emph{canonical} representation from the set of
possibilities. Unfortunately, this procedure is not standardized over the
various languages, e.g., the $3'$-rule in HGVS vs. the $5'$-rule in VCF.
Within a certain language, however, proper normalization solves the
problem of identifying \emph{equivalent} variant representations. The
implications of using non-normalized variants representations have been
reviewed in~\cite{variant_name, variant_calling, curation, litvar}.
Solutions to this problem are presented
in~\cite{unified, improved_vcf, vmc_spec, smash, plyranges, gql, ask2me, deletions}.
Often dedicated tooling~\cite{varsome, validator, mutalyzer, vde} is
needed to rigorously apply the proposed normalization procedure.
Normalized variant representations can be textually compared using
standard string matching.

\begin{figure}[ht!]
\begin{center}
\begin{tikzpicture}[yscale=-.8, minimum size=4.5mm]

\node[rectangle, fill=black!15, thick] (al0_ref0) at (0.0, 0) {\texttt{G}};
\node[rectangle, fill=black!15, thick] (al0_ref1) at (0.5, 0) {\texttt{A}};
\node[rectangle, fill=black!15, thick] (al0_ref2) at (1.0, 0) {\texttt{A}};
\node[rectangle, fill=black!15, thick] (al0_ref3) at (1.5, 0) {\texttt{T}};
\node[rectangle, fill=black!15, thick] (al0_ref4) at (2.0, 0) {\texttt{C}};
\node[rectangle, fill=black!15, thick] (al0_ref5) at (2.5, 0) {\texttt{-}};
\node[rectangle, fill=black!15, thick] (al0_ref6) at (3.0, 0) {\texttt{-}};
\node[rectangle, fill=black!15, thick] (al0_ref7) at (3.5, 0) {\texttt{G}};

\node[rectangle, fill=black!15, thick] (al0_obs0) at (0.0, 1) {\texttt{G}};
\node[rectangle, fill=black!15, thick] (al0_obs1) at (0.5, 1) {\texttt{-}};
\node[rectangle, fill=black!15, thick] (al0_obs2) at (1.0, 1) {\texttt{A}};
\node[rectangle, fill=black!15, thick] (al0_obs3) at (1.5, 1) {\texttt{T}};
\node[rectangle, fill=black!15, thick] (al0_obs4) at (2.0, 1) {\texttt{C}};
\node[rectangle, fill=black!15, thick] (al0_obs5) at (2.5, 1) {\texttt{C}};
\node[rectangle, fill=black!15, thick] (al0_obs6) at (3.0, 1) {\texttt{T}};
\node[rectangle, fill=black!15, thick] (al0_obs7) at (3.5, 1) {\texttt{G}};

\draw[semithick] (al0_ref0) -- (al0_obs0);
\draw[semithick] (al0_ref2) -- (al0_obs2);
\draw[semithick] (al0_ref3) -- (al0_obs3);
\draw[semithick] (al0_ref4) -- (al0_obs4);
\draw[semithick] (al0_ref7) -- (al0_obs7);

\node[rectangle, fill=black!15, thick] (al1_ref0) at (0.0, 2.5) {\texttt{G}};
\node[rectangle, fill=black!15, thick] (al1_ref1) at (0.5, 2.5) {\texttt{A}};
\node[rectangle, fill=black!15, thick] (al1_ref2) at (1.0, 2.5) {\texttt{A}};
\node[rectangle, fill=black!15, thick] (al1_ref3) at (1.5, 2.5) {\texttt{T}};
\node[rectangle, fill=black!15, thick] (al1_ref4) at (2.0, 2.5) {\texttt{-}};
\node[rectangle, fill=black!15, thick] (al1_ref5) at (2.5, 2.5) {\texttt{C}};
\node[rectangle, fill=black!15, thick] (al1_ref6) at (3.0, 2.5) {\texttt{-}};
\node[rectangle, fill=black!15, thick] (al1_ref7) at (3.5, 2.5) {\texttt{G}};

\node[rectangle, fill=black!15, thick] (al1_obs0) at (0.0, 3.5) {\texttt{G}};
\node[rectangle, fill=black!15, thick] (al1_obs1) at (0.5, 3.5) {\texttt{-}};
\node[rectangle, fill=black!15, thick] (al1_obs2) at (1.0, 3.5) {\texttt{A}};
\node[rectangle, fill=black!15, thick] (al1_obs3) at (1.5, 3.5) {\texttt{T}};
\node[rectangle, fill=black!15, thick] (al1_obs4) at (2.0, 3.5) {\texttt{C}};
\node[rectangle, fill=black!15, thick] (al1_obs5) at (2.5, 3.5) {\texttt{C}};
\node[rectangle, fill=black!15, thick] (al1_obs6) at (3.0, 3.5) {\texttt{T}};
\node[rectangle, fill=black!15, thick] (al1_obs7) at (3.5, 3.5) {\texttt{G}};

\draw[semithick] (al1_ref0) -- (al1_obs0);
\draw[semithick] (al1_ref2) -- (al1_obs2);
\draw[semithick] (al1_ref3) -- (al1_obs3);
\draw[semithick] (al1_ref5) -- (al1_obs5);
\draw[semithick] (al1_ref7) -- (al1_obs7);

\node[rectangle, fill=black!15, thick] (al4_ref0) at (6.0, 0) {\texttt{G}};
\node[rectangle, fill=black!15, thick] (al4_ref1) at (6.5, 0) {\texttt{A}};
\node[rectangle, fill=black!15, thick] (al4_ref2) at (7.0, 0) {\texttt{A}};
\node[rectangle, fill=black!15, thick] (al4_ref3) at (7.5, 0) {\texttt{T}};
\node[rectangle, fill=black!15, thick] (al4_ref4) at (8.0, 0) {\texttt{C}};
\node[rectangle, fill=black!15, thick] (al4_ref5) at (8.5, 0) {\texttt{-}};
\node[rectangle, fill=black!15, thick] (al4_ref6) at (9.0, 0) {\texttt{G}};

\node[rectangle, fill=black!15, thick] (al4_obs0) at (6.0, 1) {\texttt{G}};
\node[rectangle, fill=black!15, thick] (al4_obs1) at (6.5, 1) {\texttt{A}};
\node[rectangle, fill=black!15, thick] (al4_obs2) at (7.0, 1) {\texttt{-}};
\node[rectangle, fill=black!15, thick] (al4_obs3) at (7.5, 1) {\texttt{T}};
\node[rectangle, fill=black!15, thick] (al4_obs4) at (8.0, 1) {\texttt{C}};
\node[rectangle, fill=black!15, thick] (al4_obs5) at (8.5, 1) {\texttt{T}};
\node[rectangle, fill=black!15, thick] (al4_obs6) at (9.0, 1) {\texttt{G}};

\draw[semithick] (al4_ref0) -- (al4_obs0);
\draw[semithick] (al4_ref1) -- (al4_obs1);
\draw[semithick] (al4_ref3) -- (al4_obs3);
\draw[semithick] (al4_ref4) -- (al4_obs4);
\draw[semithick] (al4_ref6) -- (al4_obs6);

\node[rectangle, fill=black!15, thick] (al5_ref0) at (6.0, 2.5) {\texttt{G}};
\node[rectangle, fill=black!15, thick] (al5_ref1) at (6.5, 2.5) {\texttt{A}};
\node[rectangle, fill=black!15, thick] (al5_ref2) at (7.0, 2.5) {\texttt{A}};
\node[rectangle, fill=black!15, thick] (al5_ref3) at (7.5, 2.5) {\texttt{T}};
\node[rectangle, fill=black!15, thick] (al5_ref4) at (8.0, 2.5) {\texttt{C}};
\node[rectangle, fill=black!15, thick] (al5_ref5) at (8.5, 2.5) {\texttt{-}};
\node[rectangle, fill=black!15, thick] (al5_ref6) at (9.0, 2.5) {\texttt{G}};

\node[rectangle, fill=black!15, thick] (al5_obs0) at (6.0, 3.5) {\texttt{G}};
\node[rectangle, fill=black!15, thick] (al5_obs1) at (6.5, 3.5) {\texttt{-}};
\node[rectangle, fill=black!15, thick] (al5_obs2) at (7.0, 3.5) {\texttt{A}};
\node[rectangle, fill=black!15, thick] (al5_obs3) at (7.5, 3.5) {\texttt{T}};
\node[rectangle, fill=black!15, thick] (al5_obs4) at (8.0, 3.5) {\texttt{C}};
\node[rectangle, fill=black!15, thick] (al5_obs5) at (8.5, 3.5) {\texttt{T}};
\node[rectangle, fill=black!15, thick] (al5_obs6) at (9.0, 3.5) {\texttt{G}};

\draw[semithick] (al5_ref0) -- (al5_obs0);
\draw[semithick] (al5_ref2) -- (al5_obs2);
\draw[semithick] (al5_ref3) -- (al5_obs3);
\draw[semithick] (al5_ref4) -- (al5_obs4);
\draw[semithick] (al5_ref6) -- (al5_obs6);

\node[draw, fit=(al0_ref0) (al4_obs6)] {};
\node[draw, fit=(al1_ref0) (al5_obs6)] {};

\begin{scope}[on background layer]
\fill[orange!30] ([xshift=.4pt, yshift=.5pt]al0_ref1.north west) rectangle ([xshift=-.4pt, yshift=-.5pt]al0_obs1.south east);
\path (al0_ref1) -- (al0_obs1) node[midway] {\footnotesize $\dagger$};
\fill[orange!30] ([xshift=.4pt, yshift=.5pt]al0_ref5.north west) rectangle ([xshift=-.4pt, yshift=-.5pt]al0_obs5.south east);
\path (al0_ref5) -- (al0_obs5) node[midway] {\footnotesize $\dagger$};
\fill[blue!30]([xshift=.4pt, yshift=.5pt]al0_ref6.north west) rectangle ([xshift=-.4pt, yshift=-.5pt]al0_obs6.south east);
\path (al0_ref6) -- (al0_obs6) node[midway] {\footnotesize $\star$};

\fill[blue!30] ([xshift=.4pt, yshift=.5pt]al1_ref1.north west) rectangle ([xshift=-.4pt, yshift=-.5pt]al1_obs1.south east);
\path (al1_ref1) -- (al1_obs1) node[midway] {\footnotesize $\star$};
\fill[orange!30] ([xshift=.4pt, yshift=.5pt]al1_ref4.north west) rectangle ([xshift=-.4pt, yshift=-.5pt]al1_obs4.south east);
\path (al1_ref4) -- (al1_obs4) node[midway] {\footnotesize $\dagger$};
\fill[blue!30]([xshift=.4pt, yshift=.5pt]al1_ref6.north west) rectangle ([xshift=-.4pt, yshift=-.5pt]al1_obs6.south east);
\path (al1_ref6) -- (al1_obs6) node[midway] {\footnotesize $\star$};

\fill[orange!30] ([xshift=.4pt, yshift=.5pt]al4_ref2.north west) rectangle ([xshift=-.4pt, yshift=-.5pt]al4_obs2.south east);
\path (al4_ref2) -- (al4_obs2) node[midway] {\footnotesize $\dagger$};
\fill[blue!30] ([xshift=.4pt, yshift=.5pt]al4_ref5.north west) rectangle ([xshift=-.4pt, yshift=-.5pt]al4_obs5.south east);
\path (al4_ref5) -- (al4_obs5) node[midway] {\footnotesize $\star$};

\fill[blue!30] ([xshift=.4pt, yshift=.5pt]al5_ref1.north west) rectangle ([xshift=-.4pt, yshift=-.5pt]al5_obs1.south east);
\path (al5_ref1) -- (al5_obs1) node[midway] {\footnotesize $\star$};

\fill[blue!30] ([xshift=.4pt, yshift=.5pt]al5_ref5.north west) rectangle ([xshift=-.4pt, yshift=-.5pt]al5_obs5.south east);
\path (al5_ref5) -- (al5_obs5) node[midway] {\footnotesize $\star$};
\end{scope}

\end{tikzpicture}
\caption{The top panel shows alignments for two variants,
\texttt{GATCCTG} and \texttt{GATCTG}, with the same reference
sequence~\texttt{GAATCG}, where the \emph{changes}
(\hspace*{-1mm}\changeStar common to both, $\changeDagger$ unique for one
of them) suggest overlap. The bottom panel shows these same variants, but
now obtained through different alignments, where the changes this time
suggest that the left variant contains the right one.}\label{fig:example}
\end{center}
\end{figure}
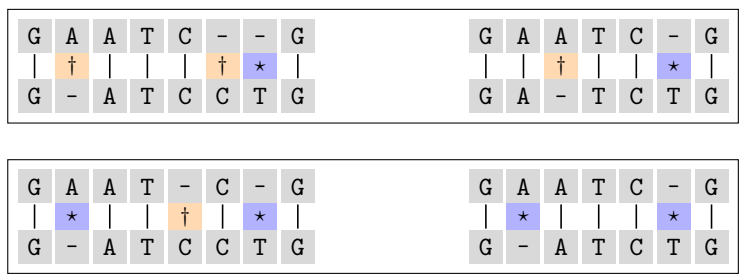

\noindent
Arguably, identification of equivalent variant representations, i.e.,
determining whether two variant descriptions result in the same observed
sequence, is currently the most interesting query in the variant domain,
as it allows for the grouping and matching of equivalent variants and
their annotations. With the advent of long read single molecule
sequencing technologies (provided by platforms such as those manufactured
by Pacific Biosciences and Oxford Nanopore), which are capable of
providing direct evidence of numerous small variants that are part of the
same haplotype, a richer set of questions arises. For example,
identification of suballeles, which is of interest in the fields of
molecular microbiology (strain typing) and pharmacogenomics (star allele
calling), can be achieved by determining whether the suballele of
interest is contained within the observed allele.

Minimal sequence level alignments, informally defined as sequences of
deletions/insertions that transform one string into another, having
shortest length possible, are used to define relations between given
variants of the same reference sequence. Figure~\ref{fig:example} shows
an example. In Section~\ref{sec:formal} we precisely define how the
relations depend on the set of all alignments between the two sequences.
In the example situation the containment relation takes precedence over
the overlap relation.

In this paper we explore the relations of variants in an exhaustive
manner. In addition to the equivalence relation, we partition the domain
of binary variant relations into Boolean relations: equivalence;
containment, i.e., either a variant is fully contained in another or a
variant fully contains another; overlap, i.e., two variants have (at
least) one common element; and disjoint, i.e., no common elements.
Because of this partitioning, exactly one of the aforementioned
relations is true for every pair of variants. For determining the
relation, we consider all (minimal) variant representations
simultaneously.

\section{Formalization}\label{sec:formal}

Formally, a \emph{variant representation} is a pair $(R, \varphi)$, where
$R$ is a \emph{string}, a finite sequence of \emph{symbols} from a
non-empty finite \emph{alphabet}, e.g.,
$\Sigma = \{\texttt{A}, \texttt{C}, \texttt{G}, \texttt{T}\}$,
called the reference sequence, and $\varphi$ is a finite set of
\emph{operations} transforming the string~$R$ into the string~$O$, the
observed sequence. The \emph{length} of a string~$S$, denoted by $|S|$,
is the number of symbols in $S$. We refer to the symbol on position~$i$
of string~$S$ as $S_i$, with $1 \le i \le |S|$. This notation is extended
in the natural way for substrings of $S$, i.e., $S_{i\ldots j}$
represents the string containing the contiguous
symbols~$S_i,\ldots, S_j$, with $1 \le i < j \le |S|$.

Note that the set of operations is dependent on the variant
representation language used. The actual problem of transforming a
reference sequence into an observed sequence is, for instance, handled
in~\cite{mutalyzer}.

The difference between the reference sequence~($R$) and the observed
sequence~($O$) is the ``actual'' variant, which is, to some extend,
independent from the original representation ($\varphi$) as we take
\emph{all} minimal representations into account. To this end we perform a
global pairwise alignment between $R$ and $O$. In contrast to the
specialized alignment methods used in, for instance, the context of short
read sequencing, we use an elementary form of alignment which is close to
a commonly used distance metric, the
\emph{Levenshtein distance}~\cite{levenshtein}.
The \emph{simple edit distance}, i.e., the Levenshtein distance without
substitutions and weighing both deletions and insertions as $1$, is
defined as the minimal number $d(R, O)$ of deletions and insertions to
transform string~$R$ into string~$O$. It can be determined by
$d(R, O) = D(|R|, |O|)$, given by the recurrence relation with
$1 \le i \le |R|$ and $1 \le j \le |O|$:

\begin{equation}\label{eq:edit}
\begin{cases}
\begin{aligned}
D(0, 0) &= 0, \\
D(i, 0)\, &= i, \\
D(0, j) &= j, \\
D(i, j) &=
\begin{cases}
D(i - 1, j - 1) & \text{if } R_i = O_j, \\
\min
\begin{cases}
D(i - 1, j) + 1, \\
D(i, j - 1) + 1 \\
\end{cases} & \text{otherwise}.
\end{cases}
\end{aligned}
\end{cases}
\end{equation}

\bigskip

\noindent
The simple edit distance is related to the
\emph{Longest Common Subsequence}~(LCS) problem~\cite{survey}:
\begin{equation}\label{eq:lcs}
D(i, j) = i + j - 2\cdot|\text{LCS}(R_{1\ldots i}, O_{1\ldots j})|.
\end{equation}

\noindent
Commonly, the recurrence relation is computed using a dynamic programming
approach by filling a matrix containing the solutions to
Equation~\ref{eq:edit} in a bottom-up fashion~\cite{string_to_string}.
Consider the computation of the simple edit distance between
$R = \texttt{CATATATCG}$ and $O = \texttt{CTTATAGCAT}$ in
Figure~\ref{fig:edit}. The simple edit distance $D(|R|, |O|) = 7$ is
given by the bottom-right element.

\begin{figure}[ht!]
\[
\begin{array}{rc|ccccccccccc}
\renewcommand*{\arraystretch}{1.2}
 & & \phantom{\circled{0}} & \phantom{\circled{1}} & \phantom{\circled{2}} & \phantom{\circled{3}} & \phantom{\circled{4}} & \phantom{\circled{5}} & \phantom{\circled{6}} & \phantom{\circled{7}} & \phantom{\circled{8}} & \phantom{\circled{9}} & \phantom{\circled{10}} \\
 & & 0 & 1 & 2 & 3 & 4 & 5 & 6 & 7 & 8 & 9 & 10 \\
 & & \texttt{.} & \texttt{C} & \texttt{T} & \texttt{T} & \texttt{A} & \texttt{T} & \texttt{A} & \texttt{G} & \texttt{C} & \texttt{A} & \texttt{T} \\
\hline
0 & \texttt{.} & \tikzmark{lcs0}0 & 1 & 2 & 3 & 4 & 5 & 6 & 7 & 8 & 9 & 10 \\
1 & \texttt{C} & 1 & \tikzmark{lcs1}\circled{0} & 1 & 2 & 3 & 4 & 5 & 6 & \circled{7} & 8 & 9 \\
2 & \texttt{A} & 2 & \tikzmark{lcs2}1 & 2 & 3 & \circled{2} & 3 & \circled{4} & 5 & 6 & \circled{7} & 8 \\
3 & \texttt{T} & 3 & 2 & \tikzmark{lcs3}\circled{1} & \tikzmark{lcs4}\circled{2} & 3 & \circled{2} & 3 & 4 & 5 & 6 & \circled{7} \\
4 & \texttt{A} & 4 & 3 & 2 & 3 & \tikzmark{lcs5}\circled{2} & 3 & \circled{2} & 3 & 4 & \circled{5} & 6 \\
5 & \texttt{T} & 5 & 4 & \circled{3} & \circled{2} & 3 & \tikzmark{lcs6}\circled{2} & 3 & 4 & 5 & 6 & \circled{5} \\
6 & \texttt{A} & 6 & 5 & 4 & 3 & \circled{2} & 3 & \tikzmark{lcs7}\circled{2} & \tikzmark{lcs8}3 & 4 & \circled{5} & 6 \\
7 & \texttt{T} & 7 & 6 & \circled{5} & \circled{4} & 3 & \circled{2} & 3 & \tikzmark{lcs9}4 & 5 & 6 & \circled{5} \\
8 & \texttt{C} & 8 & \circled{7} & 6 & 5 & 4 & 3 & 4 & 5 & \tikzmark{lcs10}\circled{4} & \tikzmark{lcs11}5 & 6 \\
9 & \texttt{G} & 9 & 8 & 7 & 6 & 5 & 4 & 5 & \circled{4} & 5 & \tikzmark{lcs12}6 & \tikzmark{lcs13}7 \\
\end{array}
\]
\begin{tikzpicture}[overlay, remember picture]
\draw[thick, draw=blue!50] (lcs0.west) -- ([xshift=2.5pt] lcs1.west);
\draw[thick, draw=blue!50] ([xshift=2.5pt] lcs1.west) -- (lcs2.west);
\draw[thick, draw=blue!50] (lcs2.west) -- (lcs3.south east);
\draw[thick, draw=blue!50] (lcs3.south east) -- (lcs4.south east);
\draw[thick, draw=blue!50] (lcs4.south east) -- (lcs7.south east);
\draw[thick, draw=blue!50] (lcs7.south east) -- (lcs8.south west);
\draw[thick, draw=blue!50] (lcs8.south west) -- (lcs9.west);
\draw[thick, draw=blue!50] (lcs9.west) -- (lcs10.south east);
\draw[thick, draw=blue!50] (lcs10.south east) -- (lcs11.south west);
\draw[thick, draw=blue!50] (lcs11.south west) -- (lcs12.south west);
\draw[thick, draw=blue!50] (lcs12.south west) -- (lcs13.south east);
\end{tikzpicture}
\caption{Computation matrix of the simple edit distance between
$R = \texttt{CATATATCG}$ and $O = \texttt{CTTATAGCAT}$. Matching symbols
are annotated with a circle. The highlighted path shows one of the
minimal alignments.}\label{fig:edit}
\end{figure}

\noindent
Informally, a representation of string~$O$ with respect to string~$R$
(the reference) is a well-defined algorithm to transform~$R$ into~$O$.
Formally, it consists of single symbol deletions/insertions (operations)
at well-defined string positions from~$R$. In the case of insertions, the
inserted symbol is also provided; for deletions this is optional.
Note that the order of the insertions matters, but deletions can be
performed in any order. An easy way to achieve all this is by indexing
the positions in~$R$ ($1,2,\ldots,|R|$), and providing each operation
with the appropriate index from this original numbering. Operations then
take place after the position mentioned, where index~$0$ is used for
insertions at the beginning. The ordering issue for insertions can also
be resolved by combining the symbols of all insertions at the same
position into one string in the desired order.

Note that many languages, like HGVS~\cite{hgvs}, can be used to
accomplish the same result. As an example, \texttt{8\_9insA} denotes the
insertion of symbol~\texttt{A} after the eighth symbol of~$R$. Likewise,
\texttt{7delT} represents a deletion of the symbol~\texttt{T} at
position~$7$ of~$R$. Together they constitute the representation
$\texttt{[7delT;8\_9insA]}$, yielding the string~$O = \texttt{CATATACAG}$
from reference~$R = \texttt{CATATATCG}$.

A minimal representation is a representation with the smallest number of
operations. Such a minimal representation uniquely corresponds to a
``path'' in the matrix from top-left to bottom-right. These paths can be
computed from the matrix by tracing back from the bottom-right element to
the top-left element while doing only orthogonal (up or left) steps for
non-matching elements if the next element has a lower value than the
current one. Vertical steps correspond to deletions, while horizontal
steps correspond to insertions.
For matching elements (circled) a diagonal step (up and left) is allowed,
keeping the current value. Note that matching elements are not recorded
in a representation, but can easily be inferred: they are exactly the
non-deleted positions. For instance,
\texttt{[2delA;3\_4insT;6\_7insG;7delT;8\_9insA;9delG;9\_10insT]}
corresponds to the highlighted minimal representation for the example in
Figure~\ref{fig:edit}. Also note that any minimal representation has the
same number of deletions, and also the same number of insertions.

The computational complexity of the simple edit distance is
$\mathcal{O}(|R| \cdot |O|)$~\cite{lcs_complex}, although many tailored
algorithms exist that have an improved bound for specific classes of
strings~\cite{survey, tour, efficient_alllcs, optimal_lcs}. In practice
this means that only a subset of the elements in the matrix needs to be
computed, in particular if only one solution (or just the distance value)
is required.

In general, the number of equivalent trace backs, called
LCS~\emph{embeddings} in~\cite{fast_alllcs, bounds}, is exponentially
bounded by $\binom{|R| + |O|}{|R|}$. We call the set of all minimal
representations~$\Phi(R, O)$ and we formalize the relations between
non-empty variants with regard to a fixed reference sequence~$R$ (we will
omit $R$ from our notation for the sake of brevity) by using their
respective~$O$~and~$P$~observed sequences as generic representations as
follows.

\begin{definition}[Equivalence]\label{def:equivalence}
Two variants~$\varphi_O$~and~$\varphi_P$ are \emph{equivalent} if and
only if $\Phi(R, O) = \Phi(R, P)$, consequently, $O = P$.
\end{definition}

\paragraph{Example:}
$R = \texttt{TTTTTT}, \quad \varphi_O = \texttt{1delT}, \quad \varphi_P = \texttt{6delT}$ \\
Here, \texttt{1delT} (HGVS omits the square brackets in case of a single
operation) and \texttt{6delT} are equivalent because their respective
sets of minimal alignments are equal. Classic normalization procedures
followed by exact string matching are sufficient to draw the same
conclusion. This does not hold for the remaining relations as they rely
on checking all combinations of all minimal alignments.

\begin{definition}[Containment]\label{def:containment}
The variant~$\varphi_O$ \emph{contains} the variant~$\varphi_P$ if and
only if $\varphi_O' \varsupsetneq \varphi_P'$ for some
$\varphi_O' \in \Phi(R, O)$ and $\varphi_P' \in \Phi(R, P)$, and
$\varphi_O$ is not equivalent to $\varphi_P$.
\end{definition}

\noindent
We find a representation within the set of minimal representations
for~$O$ that is a proper subset of a representation within the set of
minimal representations for~$P$. 

\paragraph{Example:}
$R = \texttt{TTTTTT}, \quad \varphi_O = \texttt{2\_5delinsGGG}, \quad \varphi_P = \texttt{3T>G}$ \\
\texttt{2\_5delinsGGG} (HGVS syntax for
\texttt{[2delT;3delT;4delT;5delT;5\_6insGGG]}) contains \texttt{3T>G}
(HGVS syntax for \texttt{[3delT;3\_4insG]}) and conversely by
definition, \texttt{3T>G} is contained by \texttt{2\_5delinsGGG}. The
containment relation can be easily shown by looking at
$\varphi_O' = \texttt{[1\_2insG;2delT;2\_3insG;3delT;}$ $\texttt{3\_4insG;4delT;}$ $\texttt{5delT]}$
and $\varphi_P' = \texttt{[2\_3insG;3delT]}$. All elements of
$\varphi_P'$ are found in $\varphi_O'$. Different combinations of minimal
representations for $O$ and $P$ possibly yield incomplete results:
$\varphi_O'' = \texttt{[1delT;2delT;3\_4insG;4delT;4\_5insG;5delT;5\_6insG]}$
and $\varphi_P'' = \texttt{[3delT;3\_4insG]}$, which gives a single
common element (\texttt{3\_4insG}), or even
$\varphi_P''' = \texttt{[2\_3insG;6delT]}$ without any common element
with $\varphi_O''$. However, the existence of the combination
$\varphi_O'$ and $\varphi_P'$ determines the containment relation.

\bigskip

\noindent
Notable examples of this relation can be found by comparing multiple
alleles of polymorphic simple tandem repeats, i.e., a long repeat
expansion contains all shorter ones. The variants in
Figure~\ref{fig:example} are another example of the containment relation.

\begin{definition}[Overlap]\label{def:overlap}
Two non-equivalent variants~$\varphi_O$~and~$\varphi_P$ \emph{overlap} if
and only if $\varphi_O' \cap \varphi_P' \neq \varnothing$ for some
$\varphi_O' \in \Phi(R, O)$ and $\varphi_P' \in \Phi(R, P)$ while neither
$\varphi_O$ contains $\varphi_P$ nor $\varphi_P$ contains $\varphi_O$.
\end{definition}

\noindent
A proper subset of a representation within the set of minimal
representations for~$O$ is shared with a proper subset of a
representation within the set of minimal representations for~$P$.

\paragraph{Example:}
$R = \texttt{TTTTTT}, \quad \varphi_O = \texttt{2\_4delinsGG}, \quad \varphi_P = \texttt{3T>A}$ \\
\texttt{2\_4delinsGG} has overlap with \texttt{3T>A}. A common element
(\texttt{3delT}) is easily found:
$\varphi_O' = \texttt{[1\_2insG;2delT;3delT;3\_4insG;6delT]}$ and
$\varphi_P' = \texttt{[3delT;3\_4insA]}$, however, the insertion of the
symbol \texttt{A} cannot be found in any minimal representation of $O$.
Also, the insertion of the symbol \texttt{G} (in $O$) cannot be found in
any minimal representation of $P$. In general, the makeup of the common
elements, or even the number of common elements between different
combinations of minimal representations is not constant.

\bigskip

\noindent
Polymorphic SNVs are a notable example of the overlap relation, as they
share the deleted nucleotide, but the inserted nucleotide is different by
definition.

\begin{definition}[Disjoint]\label{def:disjoint}
Two variants~$\varphi_O$~and~$\varphi_P$ are \emph{disjoint} if they are
not equivalent, are not contained in one another, and do not overlap.
\end{definition}

\noindent
None of the minimal representations of~$O$ share anything with any of the
minimal representations of~$P$.

\paragraph{Example:}
$R = \texttt{TTTTTT}, \quad \varphi_O = \texttt{2\_3insA}, \quad \varphi_P = \texttt{4\_5insA}$ \\
\texttt{2\_3insA} and \texttt{4\_5insA} are disjoint. Although both
insert the same symbol (\texttt{A}), this cannot occur at a common
position within~$R$.

\bigskip

\noindent
The properties of the Boolean relations given in
Table~\ref{tab:properties} follow directly from the aforementioned
definitions. The table is provided for completeness and future reference,
and throughout this paper we use these properties to reason about
relations.

\begin{table}[ht!]
\caption{Properties of the Boolean relations. The converse of
``contains'' is ``is contained'' and vice versa.}%
\label{tab:properties}
\begin{center}
\begin{tabular}{l | lll}
\textbf{relation} & \textbf{symmetry} & \textbf{reflexivity} & \textbf{transitivity} \\
\hline
equivalent    &  symmetric   &  reflexive    &  transitive    \\
contains      &  asymmetric  &  irreflexive  &  transitive    \\
is contained  &  asymmetric  &  irreflexive  &  transitive    \\
overlap       &  symmetric   &  irreflexive  &  intransitive  \\
disjoint      &  symmetric   &  irreflexive  &  intransitive  \\
\end{tabular}
\end{center}
\end{table}

\section{An Efficient Algorithm}\label{sec:algorithm}

The formal definitions of the Boolean relations presented in
Section~\ref{sec:formal} depend on the enumeration of all minimal variant
representations. As explained in~\cite{bounds}, the number of
representations is bounded exponentially by the length of strings~$R$
and~$O$. For large strings (such as whole human chromosomes up to
ca.~$250 \cdot 10^6$) this approach is infeasible. In this section we
present an alternative and efficient way for the computation of each of
the relations.

\paragraph{Equivalence} As follows directly from
Definition~\ref{def:equivalence}, equivalence can be computed by a string
matching over $O$ and $P$ in $\mathcal{O}(\min(|O|, |P|))$ time and
$\mathcal{O}(|O| + |P|)$ space (storing both strings). This is optimal.
Alternatively, we can compute metric~$d$ for $O$ and $P$:
$d(O, P) = 0$ if and only if $\varphi_O$ is equivalent to $\varphi_P$.

\paragraph{Containment} We observe that computing the minimal distances
is sufficient: $d(R, O) - d(R, P) = d(O, P)$ and $d(O, P) > 0$ if and
only if $\varphi_O$ contains~$\varphi_P$. Indeed, in this situation there
is a minimal path from $R$ to $O$ that passes through $P$, and both legs
are minimal too.

\paragraph{Disjoint} Again, we note that: $d(R, O) + d(R, P) = d(O, P)$
and $d(O, P) > 0$ implies $\varphi_O$
and~$\varphi_P$ are disjoint, since any any minimal paths from $O$ to $R$
and $R$ to $P$ are disjoint here. Unfortunately, the converse is not
true. Consider the counterexample $R = \texttt{CT}$, $O = \texttt{TG}$,
and $P = \texttt{GC}$. $O$ and $P$ are disjoint despite their simple edit
distances being: $d(R, O) = 2$, $d(R, P) = 2$, $d(O, P) = 2$. Their
representations, however, have no common elements:
$\Phi(R, O) = \{\texttt{[1delC;2\_3insG]}\}$ and
$\Phi(R, P) = \{\texttt{[0\_1insG;2delT]}\}$.

\bigskip

\noindent
The aforementioned distance-based approach can be efficiently computed
using any LCS distance algorithm tailored for similar strings, e.g.,
\cite{lcs_wu}. However, to separate the disjoint and overlap relations,
we need to consider all minimal representations. With the notable
exception of the naive dynamic programming approach introduced in
Section~\ref{sec:formal}, existing algorithms typically do not compute
all representations. The naive approach suffers from a
$\mathcal{O}(|R| \cdot |O|)$ space complexity rendering it infeasible
for whole human chromosomes.

\subsection{Computing all Minimal Variant Representations}\label{sec:lcs}

Here, we present an efficient algorithm to compute the relevant elements
of the recurrence relation (Equation~\ref{eq:edit}) to be able to
reconstruct \emph{all} minimal representations (alignments) within the
theoretical complexity bounds: $\mathcal{O}(|R| \cdot |O|)$ time and
using $\mathcal{O}(|R| + |O|)$ temporary space (excluding storing
the solution). In practice, because of the high similarity between $R$
and $O$ the expected run-time is linear. The output of this algorithm is
an \emph{LCS-graph}~\cite{efficient_alllcs}: a directed acyclic graph
that consists of nodes representing single symbol matches for all
LCSs. Edges connect nodes for consecutive symbols in an LCS, possibly
labeled with a representation.

We use the generic A* search algorithm~\cite{astar} which uses a
heuristic to guide the search. In general, the space requirements of
A*~search might be of concern. However, in our case, the space is
quadratically bounded by the number of elements in the matrix.
Furthermore, we demonstrate that by expanding partial solutions in a
particular order, it is possible to bound the space requirements
linearly: $\mathcal{O}(|R| + |O|)$.

\bigskip

\noindent
We introduce the \emph{admissible heuristic}:
\begin{equation}\label{eq:heuristic}
h(R, O, i, j) = \left\lvert\left(|R| - i\right) - \left(|O| - j\right)\right\rvert.
\end{equation}

\noindent
The heuristic~$h$ represents a best-case guess for the minimal distance
from the current element~$(i, j)$ to the bottom-right element of the
matrix (hoping to match as many symbols as possible). A* minimizes the
total cost function for each solution:
\begin{equation}\label{eq:astar}
f(R, O, i, j) = D(i, j) + h(R, O, i, j),
\end{equation}

\noindent
by taking into account the actual cost to reach element~$(i, j)$, given
by $D(i, j)$ (see Equation~\ref{eq:edit}), and the estimated minimal
cost~$h$.
A*~search iteratively expands partial solutions, also called the
\emph{frontier}, based on the lowest $f$-value until the target element
is expanded. In our case the progression of $f$-values is determined by
the heuristic value of the first element $h(R, O, 0, 0) = ||R| - |O||$,
increasing with steps of~$2$, as $D$ increases by $1$ for each orthogonal
step and the heuristic changes with either $+1$ or $-1$ for each
orthogonal step. Diagonal steps, i.e., matching symbols, do not incur a
change in $f$-value. This results in a constant parity for the
$f$-values. The simple edit distance is given by the $f$-value of the
target element~$(|R|, |O|)$. Constructing all minimal variant
representations is analogous to the naive approach detailed in
Section~\ref{sec:formal}.

In typical A* implementations, the frontier is implemented as a priority
queue. In our case, we observe that we can keep track of the elements in
the frontier by describing a ``convex'' shape in the matrix. We use two
arrays~$\mathit{rows}$ and~$\mathit{cols}$ that store the right-most
element for a given column and the bottom-most element for a given row
respectively.

\begin{figure}[ht!]
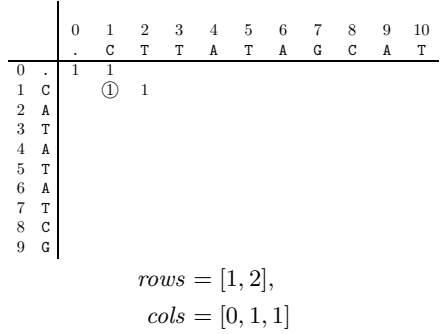
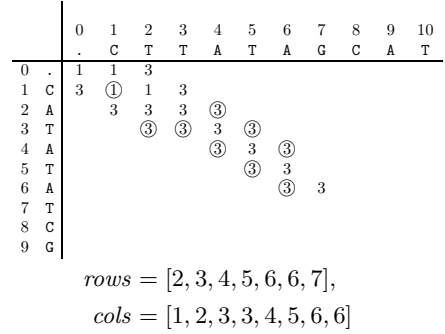
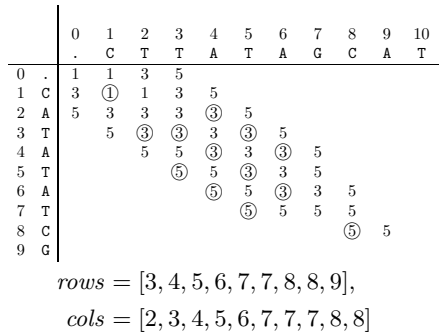
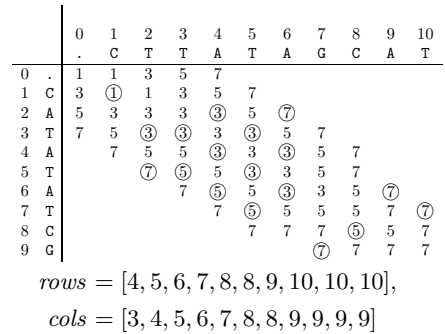

\subfloat[Expanded elements for $f = 1$.]{
\begin{minipage}[c][1\width]{
0.45\textwidth}
\centering
\[
\scalemath{0.62}{
\begin{array}{rc|ccccccccccc}
\renewcommand*{\arraystretch}{1.2}
 & & \phantom{\circled{0}} & \phantom{\circled{1}} & \phantom{\circled{2}} & \phantom{\circled{3}} & \phantom{\circled{4}} & \phantom{\circled{5}} & \phantom{\circled{6}} & \phantom{\circled{7}} & \phantom{\circled{8}} & \phantom{\circled{9}} & \phantom{\circled{10}} \\
 & & 0 & 1 & 2 & 3 & 4 & 5 & 6 & 7 & 8 & 9 & 10 \\
 & & \texttt{.} & \texttt{C} & \texttt{T} & \texttt{T} & \texttt{A} & \texttt{T} & \texttt{A} & \texttt{G} & \texttt{C} & \texttt{A} & \texttt{T} \\
\hline
0 & \texttt{.} & 1 & 1 &  &  &  &  &  &  &  &  &  \\
1 & \texttt{C} &  & \circled{1} & 1 &  &  &  &  &  &  &  &  \\
2 & \texttt{A} &  &  &  &  &  &  &  &  &  &  &  \\
3 & \texttt{T} &  &  &  &  &  &  &  &  &  &  &  \\
4 & \texttt{A} &  &  &  &  &  &  &  &  &  &  &  \\
5 & \texttt{T} &  &  &  &  &  &  &  &  &  &  &  \\
6 & \texttt{A} &  &  &  &  &  &  &  &  &  &  &  \\
7 & \texttt{T} &  &  &  &  &  &  &  &  &  &  &  \\
8 & \texttt{C} &  &  &  &  &  &  &  &  &  &  &  \\
9 & \texttt{G} &  &  &  &  &  &  &  &  &  &  &  \\
\end{array}
}
\]
\vspace{-2em}
{\small
\begin{align*}
\mathit{rows} &= [1, 2], \\
\mathit{cols} &= [0, 1, 1]
\end{align*}}%
\end{minipage}}
\hfill
\subfloat[Expanded elements for $f = 3$.]{
\begin{minipage}[c][1\width]{
0.45\textwidth}
\centering
\[
\scalemath{0.62}{
\begin{array}{rc|ccccccccccc}
\renewcommand*{\arraystretch}{1.2}
 & & \phantom{\circled{0}} & \phantom{\circled{1}} & \phantom{\circled{2}} & \phantom{\circled{3}} & \phantom{\circled{4}} & \phantom{\circled{5}} & \phantom{\circled{6}} & \phantom{\circled{7}} & \phantom{\circled{8}} & \phantom{\circled{9}} & \phantom{\circled{10}} \\
 & & 0 & 1 & 2 & 3 & 4 & 5 & 6 & 7 & 8 & 9 & 10 \\
 & & \texttt{.} & \texttt{C} & \texttt{T} & \texttt{T} & \texttt{A} & \texttt{T} & \texttt{A} & \texttt{G} & \texttt{C} & \texttt{A} & \texttt{T} \\
\hline
0 & \texttt{.} & 1 & 1 & 3 &  &  &  &  &  &  &  &  \\
1 & \texttt{C} & 3 & \circled{1} & 1 & 3 &  &  &  &  &  &  &  \\
2 & \texttt{A} &  & 3 & 3 & 3 & \circled{3} &  &  &  &  &  &  \\
3 & \texttt{T} &  &  & \circled{3} & \circled{3} & 3 & \circled{3} &  &  &  &  &  \\
4 & \texttt{A} &  &  &  &  & \circled{3} & 3 & \circled{3} &  &  &  &  \\
5 & \texttt{T} &  &  &  &  &  & \circled{3} & 3 &  &  &  &  \\
6 & \texttt{A} &  &  &  &  &  &  & \circled{3} & 3 &  &  &  \\
7 & \texttt{T} &  &  &  &  &  &  &  &  &  &  &  \\
8 & \texttt{C} &  &  &  &  &  &  &  &  &  &  &  \\
9 & \texttt{G} &  &  &  &  &  &  &  &  &  &  &  \\
\end{array}
}
\]
\vspace{-2em}
{\small
\begin{align*}
\mathit{rows} &= [2, 3, 4, 5, 6, 6, 7], \\
\mathit{cols} &= [1, 2, 3, 3, 4, 5, 6, 6]
\end{align*}}%
\end{minipage}}

\subfloat[Expanded elements for $f = 5$.]{
\begin{minipage}[c][1\width]{
0.45\textwidth}
\centering
\[
\scalemath{0.62}{
\begin{array}{rc|ccccccccccc}
\renewcommand*{\arraystretch}{1.2}
 & & \phantom{\circled{0}} & \phantom{\circled{1}} & \phantom{\circled{2}} & \phantom{\circled{3}} & \phantom{\circled{4}} & \phantom{\circled{5}} & \phantom{\circled{6}} & \phantom{\circled{7}} & \phantom{\circled{8}} & \phantom{\circled{9}} & \phantom{\circled{10}} \\
 & & 0 & 1 & 2 & 3 & 4 & 5 & 6 & 7 & 8 & 9 & 10 \\
 & & \texttt{.} & \texttt{C} & \texttt{T} & \texttt{T} & \texttt{A} & \texttt{T} & \texttt{A} & \texttt{G} & \texttt{C} & \texttt{A} & \texttt{T} \\
\hline
0 & \texttt{.} & 1 & 1 & 3 & 5 &  &  &  &  &  &  &  \\
1 & \texttt{C} & 3 & \circled{1} & 1 & 3 & 5 &  &  &  &  &  &  \\
2 & \texttt{A} & 5 & 3 & 3 & 3 & \circled{3} & 5 &  &  &  &  &  \\
3 & \texttt{T} &  & 5 & \circled{3} & \circled{3} & 3 & \circled{3} & 5 &  &  &  &  \\
4 & \texttt{A} &  &  & 5 & 5 & \circled{3} & 3 & \circled{3} & 5 &  &  &  \\
5 & \texttt{T} &  &  &  & \circled{5} & 5 & \circled{3} & 3 & 5 &  &  &  \\
6 & \texttt{A} &  &  &  &  & \circled{5} & 5 & \circled{3} & 3 & 5 &  &  \\
7 & \texttt{T} &  &  &  &  &  & \circled{5} & 5 & 5 & 5 &  &  \\
8 & \texttt{C} &  &  &  &  &  &  &  &  & \circled{5} & 5 &  \\
9 & \texttt{G} &  &  &  &  &  &  &  &  &  &  &  \\
\end{array}
}
\]
\vspace{-2em}
{\small
\begin{align*}
\mathit{rows} &= [3, 4, 5, 6, 7, 7, 8, 8, 9], \\
\mathit{cols} &= [2, 3, 4, 5, 6, 7, 7, 7, 8, 8]
\end{align*}}%
\end{minipage}}
\hfill
\subfloat[Expanded elements for $f = 7$.]{
\begin{minipage}[c][1\width]{
0.45\textwidth}
\centering
\[
\scalemath{0.62}{
\begin{array}{rc|ccccccccccc}
\renewcommand*{\arraystretch}{1.2}
 & & \phantom{\circled{0}} & \phantom{\circled{1}} & \phantom{\circled{2}} & \phantom{\circled{3}} & \phantom{\circled{4}} & \phantom{\circled{5}} & \phantom{\circled{6}} & \phantom{\circled{7}} & \phantom{\circled{8}} & \phantom{\circled{9}} & \phantom{\circled{10}} \\
 & & 0 & 1 & 2 & 3 & 4 & 5 & 6 & 7 & 8 & 9 & 10 \\
 & & \texttt{.} & \texttt{C} & \texttt{T} & \texttt{T} & \texttt{A} & \texttt{T} & \texttt{A} & \texttt{G} & \texttt{C} & \texttt{A} & \texttt{T} \\
\hline
0 & \texttt{.} & 1 & 1 & 3 & 5 & 7 &  &  &  &  &  &  \\
1 & \texttt{C} & 3 & \circled{1} & 1 & 3 & 5 & 7 &  &  &  &  &  \\
2 & \texttt{A} & 5 & 3 & 3 & 3 & \circled{3} & 5 & \circled{7} &  &  &  &  \\
3 & \texttt{T} & 7 & 5 & \circled{3} & \circled{3} & 3 & \circled{3} & 5 & 7 &  &  &  \\
4 & \texttt{A} &  & 7 & 5 & 5 & \circled{3} & 3 & \circled{3} & 5 & 7 &  &  \\
5 & \texttt{T} &  &  & \circled{7} & \circled{5} & 5 & \circled{3} & 3 & 5 & 7 &  &  \\
6 & \texttt{A} &  &  &  & 7 & \circled{5} & 5 & \circled{3} & 3 & 5 & \circled{7} &  \\
7 & \texttt{T} &  &  &  &  & 7 & \circled{5} & 5 & 5 & 5 & 7 & \circled{7} \\
8 & \texttt{C} &  &  &  &  &  & 7 & 7 & 7 & \circled{5} & 5 & 7 \\
9 & \texttt{G} &  &  &  &  &  &  &  & \circled{7} & 7 & 7 & 7 \\
\end{array}
}
\]
\vspace{-2em}
{\small
\begin{align*}
\mathit{rows} &= [4, 5, 6, 7, 8, 8, 9, 10, 10, 10], \\
\mathit{cols} &= [3, 4, 5, 6, 7, 8, 8, 9, 9, 9, 9]
\end{align*}}%
\end{minipage}}
\caption{Computing the elements of Equation~\ref{eq:astar} to efficiently
reconstruct the set of all minimal variant representations.}%
\label{fig:astar}
\end{figure}

In Figure~\ref{fig:astar} we present the progression of the expansion of
the matrix elements for the example introduced in Figure~\ref{fig:edit}:
$R = \texttt{CATATATCG}$ and $O = \texttt{CTTATAGCAT}$. We use
$\mathcal{O}(|R| + |O|)$~space (excluding the output) and we expand at
most $\mathcal{O}(|R| \cdot |O|)$~elements.

\noindent
The non-filled elements are not part of any minimal representation as
they would have a greater $f$-value than the bottom-right element. The
circled elements are needed to create the LCS-graph and therefore stored.
The remaining elements are expanded, but not stored. For each circled
element we determine its place in an LCS (and level in the LCS-graph) by:
\begin{equation}\label{eq:level}
\left\lfloor\frac{i + j - D(i, j)}{2}\right\rfloor.
\end{equation}

\begin{figure}[ht!]
\begin{center}
\resizebox{\textwidth}{!}
{
\begin{tikzpicture}
\node[align=center] at (2, 3.5) {$\ell = 0$};
\node[align=center] at (5, 3.5) {$\ell = 1$};
\node[align=center] at (8, 3.5) {$\ell = 2$};
\node[align=center] at (11, 3.5) {$\ell = 3$};
\node[align=center] at (14, 3.5) {$\ell = 4$};
\node[align=center] at (17, 3.5) {$\ell = 5$};

\node[align=center, draw, circle, fill] (00) at (0, 0) {};

\node[align=center, draw, ellipse, thick, minimum width=4.8em] (11) at (2, 0) {\texttt{C}\\ $(1, 1)$};

\node[align=center, draw, ellipse, thick, minimum width=4.8em] (32) at (5, -2) {\texttt{T}\\ $(3, 2)$};
\node[align=center, draw, ellipse, thick, minimum width=4.8em] (33) at (5, 0) {\texttt{T}\\ $(3, 3)$};
\node[align=center, draw, ellipse, thick, minimum width=4.8em] (24) at (5, 2) {\texttt{A}\\ $(2, 4)$};

\node[align=center, draw, ellipse, thick, minimum width=4.8em] (53) at (8, -2) {\texttt{T}\\ $(5, 3)$};
\node[align=center, draw, ellipse, thick, minimum width=4.8em] (44) at (8, 0) {\texttt{A}\\ $(4, 4)$};
\node[align=center, draw, ellipse, thick, minimum width=4.8em] (35) at (8, 2) {\texttt{T}\\ $(3, 5)$};

\node[align=center, draw, ellipse, thick, minimum width=4.8em] (64) at (11, -2) {\texttt{A}\\ $(6, 4)$};
\node[align=center, draw, ellipse, thick, minimum width=4.8em] (55) at (11, 0) {\texttt{T}\\ $(5, 5)$};
\node[align=center, draw, ellipse, thick, minimum width=4.8em] (46) at (11, 2) {\texttt{A}\\ $(4, 6)$};

\node[align=center, draw, ellipse, thick, minimum width=4.8em] (75) at (14, -2) {\texttt{T}\\ $(7, 5)$};
\node[align=center, draw, ellipse, thick, minimum width=4.8em] (66) at (14, 0) {\texttt{A}\\ $(6, 6)$};
\node[align=center, draw, ellipse, thick, minimum width=4.8em] (69) at (14, 2) {\texttt{A}\\ $(6, 9)$};

\node[align=center, draw, ellipse, thick, minimum width=4.8em] (88) at (17, -2) {\texttt{C}\\ $(8, 8)$};
\node[align=center, draw, ellipse, thick, minimum width=4.8em] (97) at (17, 0) {\texttt{G}\\ $(9, 7)$};
\node[align=center, draw, ellipse, thick, minimum width=4.8em] (710) at (17, 2) {\texttt{T}\\ $(7, 10)$};

\node[align=center, draw, circle, fill] (1011) at (20, 0) {};

\draw[-{Latex[width=.5em]}, semithick] (00) -- (11);

\draw[-{Latex[width=.5em]}, semithick] (11) -- (24) node[above, left, midway] {\footnotesize \texttt{1\_2insTT}};
\draw[-{Latex[width=.5em]}, semithick] (11) -- (33) node[above, midway] {\footnotesize \texttt{2A>T}};
\draw[-{Latex[width=.5em]}, semithick] (11) -- (32) node[below, left, midway] {\footnotesize \texttt{2delA}};

\draw[-{Latex[width=.5em]}, semithick] (24) -- (35);
\draw[-{Latex[width=.5em]}, semithick] (33) -- (44);
\draw[-{Latex[width=.5em]}, semithick] (32) -- (53) node[above, midway] {\footnotesize \texttt{4delA}};
\draw[-{Latex[width=.5em]}, semithick] (32) -- (44) node[above, left, midway] {\footnotesize \texttt{3\_4insT}};

\draw[-{Latex[width=.5em]}, semithick] (35) -- (46);
\draw[-{Latex[width=.5em]}, semithick] (44) -- (55);
\draw[-{Latex[width=.5em]}, semithick] (53) -- (64);

\draw[-{Latex[width=.5em]}, semithick] (46) -- (69) node[above, midway] {\footnotesize \texttt{5delinsGC}};
\draw[-{Latex[width=.5em]}, semithick] (55) -- (66);
\draw[-{Latex[width=.5em]}, semithick] (55) -- (69) node[above, left, midway] {\footnotesize \texttt{5\_6insAGC}};
\draw[-{Latex[width=.5em]}, semithick] (64) -- (75);

\draw[-{Latex[width=.5em]}, semithick] (66) -- (710) node[above, left, midway] {\footnotesize \texttt{6\_7insGCA}};
\draw[-{Latex[width=.5em]}, semithick] (66) -- (88) node[below, near end, xshift=-.5em] {\footnotesize \texttt{7T>G}};
\draw[-{Latex[width=.5em]}, semithick] (66) -- (97) node[above, midway] {\footnotesize \texttt{7\_8delTC}};
\draw[-{Latex[width=.5em]}, semithick] (69) -- (710);
\draw[-{Latex[width=.5em]}, semithick] (75) -- (88) node[below, midway] {\footnotesize \texttt{7\_8insAG}};
\draw[-{Latex[width=.5em]}, semithick] (75) -- (97) node[above, near end, xshift=-.5em] {\footnotesize \texttt{8C>A}};

\draw[-{Latex[width=.5em]}, semithick] (710) -- (1011) node[above, right, midway] {\footnotesize \texttt{8\_9delCG}};
\draw[-{Latex[width=.5em]}, semithick] (88) -- (1011) node[below, right, midway] {\footnotesize \texttt{9delinsAT}};
\draw[-{Latex[width=.5em]}, semithick] (97) -- (1011) node[above, midway, xshift=-.5em] {\footnotesize \texttt{9\_10insCAT}};

\end{tikzpicture}}
\end{center}
\caption{The LCS-graph for $R = \texttt{CATATATCG}$ and
$O = \texttt{CTTATAGCAT}$. The coordinates refer to the coordinates of
the matching symbols in Figure~\ref{fig:astar}. Unlabeled edges
indicate consecutive matches and do not contribute to the set of
elements of all minimal variant representations.}\label{fig:graph}
\end{figure}
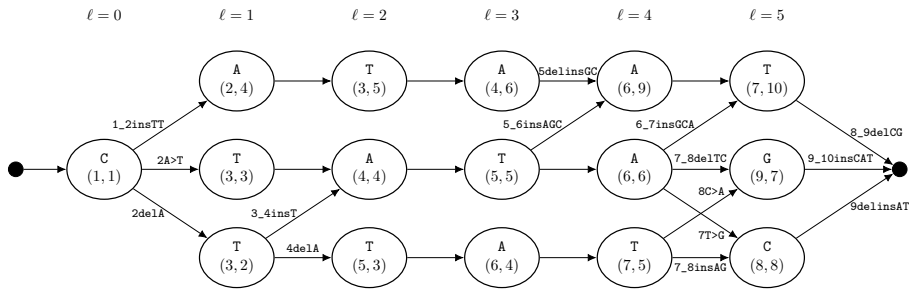

\noindent
This allows us to construct the LCS-graph efficiently. The LCS-graph for
the example in Figure~\ref{fig:astar} is given in Figure~\ref{fig:graph}.
The nodes in the LCS-graph are ordered by their position in the LCS. To
construct the variant representations, edges are added for each
node~$(i, j)$ on level~$\ell$ (determined by Equation~\ref{eq:level}) to
each node~$(i', j')$ on level~$\ell + 1$ if $\mathit{i'} > \mathit{i}$
and $\mathit{j'} > \mathit{j}$.
For instance, there is an edge from node~$(3, 2)$ on level~$1$ to
node~$(5, 3)$ on level~$2$ (\texttt{4delA}). Not all circled elements end
up in the LCS-graph as some do not lie on an optimal path, e.g.
\texttt{T} at $(5, 2)$. These elements may be represented as nodes in the
LCS-graph. For these nodes there is no path to the sink node.
Alternatively, constructing the LCS-graph from the sink node to the
source node, these elements are avoided.

We define $\Psi(R,O)$ as the set of all elements that occur in minimal
representations from $\Phi(R,O)$. To distinguish between the relations
disjoint and overlap, it is sufficient to determine whether the two
sets~$\Psi(R, O)$ and~$\Psi(R, P)$ are disjoint. Note that the number of
elements in each set is bounded quadratically as opposed to enumerating
all, exponentially bounded, minimal representations. Some practical
implementation enhancements can also be applied, notably, reducing the
number of elements to be added to the set by taking (partially)
overlapping edges in the LCS-graph into account. For small alphabets,
e.g., DNA~nucleotides, an efficient bit string can be used in lieu of a
proper set implementation.

\subsection{Maximal Influence Interval}\label{sec:influence}

Given any pair of variants (within the context of the same reference
sequence), it is likely that their relation is disjoint purely based on
their often distant positions in the reference sequence. These disjoint
relations can be determined efficiently at the cost of some
pre-computation for individual variants (n.b. not pairs of variants).

For each variant the \emph{maximal influence interval} is defined as the
interval given by the lowest row index for a deletion or an insertion
in an optimal path in~$D$ and the highest row index for a deletion or an
insertion in an optimal path in~$D$.
This interval gives the extreme bounds, as positions in the reference
sequence, of possible changes due to this variant. A pair of variants can
only be non-disjoint when their maximal influence intervals intersect.
The pre-computing of the maximal influence intervals of individual
variants is specifically worthwhile in the context of repeated querying,
e.g., a (locus specific) database and VCF annotation.

For example, given a fixed reference $R = \texttt{TCCCTTTA}$.
The variants~$\varphi_O = \texttt{3C>A}$ ($O = \texttt{TCACTTTA}$) with
maximal influence interval $[2, 5)$ and $\varphi_P = \texttt{6T>G}$
($P = \texttt{TCCCTGTA}$) with maximal influence interval $[5, 8)$ are
disjoint based on the empty intersection of their maximal influence
intervals. The variants~$\varphi_O$ and
$\varphi_{P'} = \texttt{[4del;5\_6insC]}$ ($P' = \texttt{TCCTCTTA}$) with
maximal influence interval $[2, 8)$ have intersecting intervals, and
indeed the variants overlap.
In contrast, the variants~$\varphi_O$ and
$\varphi_{P''} = \texttt{2\_3insT}$ ($P'' = \texttt{TCTCCTTTA}$) with
maximal influence interval $[2, 2)$ also have intersecting intervals, but
the variants are ultimately disjoint.

\section{Experiments}\label{sec:experiments}

To obtain an intuition of the impact of the proposed approach, we
analyzed the well-studied \textit{CFTR}~gene~(NG\_016465.4 with
$257,\!188$bp), that provides instructions for making the cystic
fibrosis transmembrane conductance regulator protein.

In dbSNP (build~154)~\cite{dbsnp} there are $62,\!215$~interpretable
variants for the \textit{CFTR}~gene which lead to
$1,\!935,\!322,\!005$~pairs of variants to analyze. Using the
method described in Section~\ref{sec:influence}, only
$92,\!251$~eligible pairs of variants with a potential non-disjoint
relation remain.

\begin{table}[ht!]
\caption{Relation counts for the pairwise comparison of variants in the
\textit{CFTR}~gene. The counts are given based on the upper triangular
matrix, so the converse relations are not included.}\label{tab:relcounts}
\begin{center}
\begin{tabular}{l | r}
\textbf{relation} & \textbf{count}  \\
\hline
equivalent   &        $0$  \\
contains     &  $5,\!491$  \\
is contained &  $4,\!629$  \\
overlap      & $37,\!690$  \\
disjoint     & $44,\!441$  \\
\end{tabular}
\end{center}
\end{table}

\noindent
When the algebra is applied to the remaining pairs, we obtain the results
in Table~\ref{tab:relcounts}. We observe that (as expected) there are no
equivalent variants for \textit{CFTR} in dbSNP, indicating a correct
application of standard normalization techniques. Beyond equivalence,
there are $10,\!120$~containment relations (either contains or is
contained), $37,\!690$~pairs have some form of overlap, and
$44,\!441$~pairs are disjoint.

\begin{figure}[ht!]
\begin{center}
\begin{tikzpicture}
\begin{axis}[%
height=7cm,
width=10cm,
axis y line*=left,
axis x line*=bottom,
ticklabel style={font=\small},
xlabel=length of maximal influence interval,
ylabel=average number of relations
]
\addplot[blue!50,fill=blue!30,only marks] table [%
x=length,
y=count]{\influencelength};
\end{axis}
\end{tikzpicture}
\caption{Scatterplot of the average number of non-disjoint relations for
all variants in \textit{CFTR} with a certain maximal influence interval
length.}\label{fig:influence_scatter}
\end{center}
\end{figure}
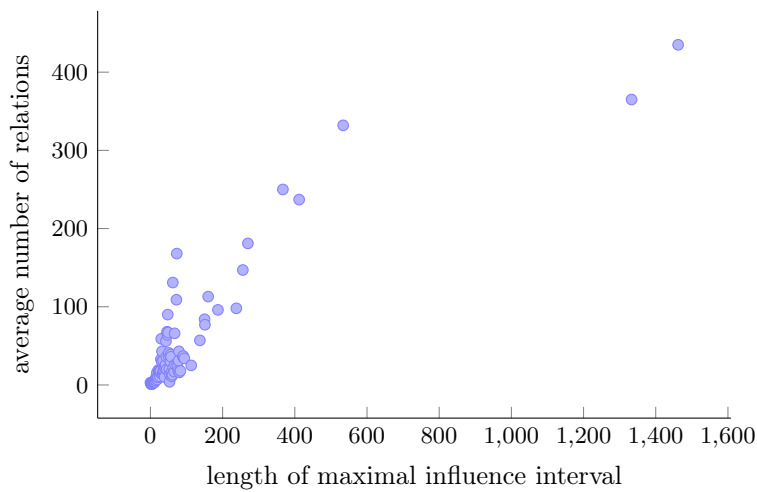

Zooming in to individual variant level (as opposed to pairs), we find
that $16,\!939$~variants are disjoint with all other variants based on
their maximal influence intervals alone and $45,\!276$~variants are
potentially involved in a non-disjoint relation with another variant.
After determining the relations, $16,\!814$~variants also turn out to be
disjoint with all other variants. In total, $33,\!753$~variants are
disjoint with all other variants. The remaining $28,\!462$~variants have
a non-disjoint relation to some other variant(s).

In Table~\ref{tab:examples}, we see a selection of variants in
\textit{CFTR} that, at first sight, have a counter-intuitive relation
with another variant. For pair~$1$, the left hand side (LHS) variant
contains the right hand side (RHS) variant because the former can be
left-justified to \texttt{11402\_11406del} (HGVS syntax for the
deletion of the symbols on positions $11,\!402,\ldots,11,\!406$) to
incorporate the deletion of region $11,\!402$ to $11,\!403$. For
pair~$2$, the containment is less obvious, the LHS needs to be rewritten
to \texttt{[151240\_151241insTATA;151270\_151271insCA]} to make this
containment relation intuitively clear. For pair~$3$, the LHS can be
written as \texttt{[151242\_151243del;151271\_151278del]} to make the
overlap relation between the two variants clear. For pair~$4$,
left-justification of the LHS to \texttt{112270\_112271insCTCTCTC} and
rewriting the RHS to \texttt{[112269\_112270insCC;} \texttt{112270\_112271insCTCT]}
makes the overlap relation obvious. Finally, we can see from both
pair~$2$ and~$3$ that in practice, variants that are reported to be well
separated, still may have something in common.

\begin{table}[hb!]
\caption{Examples of non-trivial relations between variants in
\textit{CFTR}. The variants are described using the HGVS nomenclature
with respect to reference sequence NG\_016465.4 using the genomic~(g.)
coordinate system.}\label{tab:examples}
\begin{center}
\begin{tabular}{l|l|l|l}
\textbf{pair} & \textbf{LHS variant} & \textbf{relation} & \textbf{RHS variant} \\
\hline
$1$ & \texttt{11404\_11408del}          & contains & \texttt{11402\_11403del}         \\
$2$ & \texttt{151270\_151271insTATACA}  & contains & \texttt{151240\_151241insAT}     \\
$3$ & \texttt{151271\_151280del}        & overlap  & \texttt{151240\_151255del}       \\
$4$ & \texttt{112274\_112275insCTCTCTC} & overlap  & \texttt{112269\_112270insCCTCTC} \\
\end{tabular}
\end{center}
\end{table}

\noindent
The ratio between the length of the maximal influence interval and the
number of non-disjoint relations a variant has on average is shown in
Figure~\ref{fig:influence_scatter}. The length of the maximal influence
interval correlates strongly with the number of relations of a variant as
expected. The variants with the largest maximal influence interval
lengths ($>\!150$) all happen to be large deletions, e.g.,
\texttt{203907\_204783del} contains $31$~smaller deletions and overlaps
with $404$~variants.

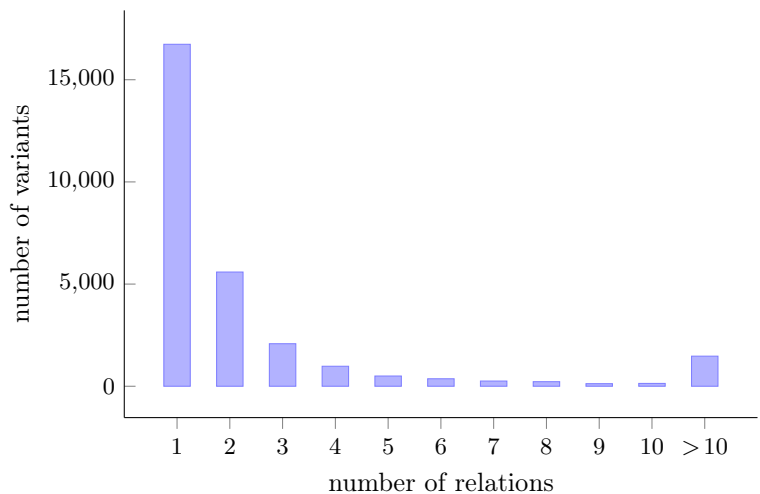
\begin{figure}[ht!]
\begin{center}
\begin{tikzpicture}
\begin{axis}[%
height=7cm,
width=10cm,
xlabel=number of relations,
axis y line*=left,
axis x line*=bottom,
ticklabel style={font=\small},
xticklabels={$1$, $2$, $3$, $4$, $5$, $6$, $7$, $8$, $9$, $10$, $>\!10$},
xtick={1,...,11},
yticklabel style={/pgf/number format/fixed,},
scaled y ticks=false,
ylabel=number of variants,
ybar]
\addplot[blue!50,fill=blue!30] table[%
x=relations,
y=count] {\relationscount};
\end{axis}
\end{tikzpicture}
\caption{The distribution of the number of non-disjoint relations per
variant. The long tail of counts of $11$ and above are aggregated. The
most relations a single variant has is $435$.}\label{fig:rel_count}
\end{center}
\end{figure}

\noindent
The distribution of the number of non-disjoint relations per variant is
shown in Figure~\ref{fig:rel_count}. More than half of all
variants ($16,\!735$) have a single non-trivial relation with another
variant, the remaining $11,\!727$~variants have a non-trivial relation
with multiple variants. The distributions for both overlap and inclusion
relations, are nearly identical.

\section{Discussion}\label{sec:discussion}

Higher-order operations like SNVs, multi-nucleotide variants,
duplications, transpositions, and inversions, can all be represented as
combinations of deletions and insertions. In practice, this view aligns
well with the expected outcomes, e.g., an SNV can be contained within a
larger deletion/insertion. Arguably, inversions are the exception, as
their distance represented as a deletion/insertion might not reflect
their true nature. This can be considered a limitation of the approach.

The relation between a pair of variants is only well-defined when both
variants are described in the context of the same reference sequence. In
general, we can extend the definitions to include variants on different
reference sequences, the natural interpretation of which would be to
consider two variants on different reference sequences to be disjoint,
e.g., a variant on human chromosome~1 has nothing in common with a
variant on human chromosome~2. This interpretation is sensible as long as
the reference sequences are unrelated. In practice however, many
reference sequences are actually referring to the same (or a strongly
related) genetic locus, e.g., genes on chromosomes, different transcripts
for the same gene and chromosomes in different reference genomes.
Arguably, variants described in the context of these reference sequences
could be seen as having potentially a non-disjoint relation. To properly
compare these variants on sequence level, the differences between the
reference sequences should also be taken into account.

Structural variants are often reported in non-exact manner, i.e., not
sequence level precise. These representations are unsuitable for our
method. Even if an exact structural variant representation is given, it
unlikely to yield meaningful results; as the exact positions are not the
same across samples. Instead, e.g., gene copies can be analyzed by the
algebra when they are provided individually.

The choice of relations presented here follows the ones from set theory,
commonly used in a wide range of domains. For some specific domains more
refined relations exists as well, e.g., for intervals the relations:
``starts with'', ``ends with'' and ``is directly adjacent'' are useful
extensions~\cite{interval}. The set of relations could be further
partitioned using these, or other, refinements.

Unfortunately, the set of relations (see Table~\ref{tab:properties}) does
not contain a relation that implies an ordering of variants, i.e.,
$\Phi(R, O) \le \Phi(R, P)$. A partial order of variants would require a
relation with the properties: reflexive, antisymmetric and transitive.
Sorting variants or storing variants in a particular order in a database
(indexing) is meaningless in the context of this algebra. The interval
ordering based on the pre-computed maximal influence intervals described
in Section~\ref{sec:influence} mitigates this problem.

\subsection{Characterization of Overlap}

The actual makeup of the common changes between two variants is never
computed. For all relations except the overlap relation the common
changes can be trivially given: none for disjoint variants, either of the
variants for equivalence, and the ``smaller'' variant for containment,
i.e., the one that is contained within the other. This leaves, however,
the overlapping variants. In general, there are many different sets of
common changes between overlapping variants, some of which, especially
the larger ones, may be more (biologically) relevant than others. The
algorithm described in Section~\ref{sec:algorithm} determines whether
there is at least one common change. Computing the maximal size of the
overlap requires enumerating an exponential number of possible
alignments, which is infeasible for all but extremely short sequences.

\subsection{General Normalization}\label{sec:general}

The current practice of normalizing variant representations is
sufficiently powerful to cater for the equivalence relation (also
illustrated in Section~\ref{sec:experiments}). Determining other
relations is, in general, impossible when given a single normalized
representation. Even SNVs, often regarded as trivially normalized, are
problematic when querying for containment. Consider
reference~$R = \texttt{CACAT}$ and the SNV~\texttt{3C>T} to obtain the
observed sequence~$O = \texttt{CATAT}$. In the classical sense no
normalization is necessary. When we consider a second
variant~\texttt{3\_4insT} (\texttt{CACTAT}) we might draw the conclusion
that this insertion is contained within the SNV based on the normalized
position. A possible third variant~\texttt{2\_3insT} (\texttt{CATCAT})
has the same relation, but is less trivially found. When substrings
adjacent to the variant match subsequences of the deleted or inserted
string, the number of alignments increases exponentially, so regardless
of which normalization procedure is used, however sophisticated, counter
examples like this can always be constructed. Therefore procedures that
rely on normalization will, in general, lead to wrong conclusions and
cannot be employed to determine relations between variants.

Within the domain specific languages for variant representations
different normalization schemes are used, where arbitrary choices
influence the normalized representation, e.g., the $3'$ and $5'$-rules.
From the alignment matrix~$D$, it is also possible to choose a canonical
path that represents a normalized representation.
Sensible choices are either a bottom-most or top-most path. This
corresponds to favoring either deletions over insertions at the beginning
of a variant (or vice versa). Note that for all minimal variant
descriptions in any of the domain specific languages, corresponding
alignments can be found. It could be worthwhile to investigate whether a
comprehensive set of deterministic rules exist to find these alignments
as this can be used in the formalization of these languages.

\subsection{Non-minimal Variant Representations}\label{sec:nonminimal}

So far, we assumed that all variant representations are minimal with
regard to Equation~\ref{eq:edit}. In practice, this is not always the
case, nor is it necessary for our approach to work as the only constraint
on the variant representation is its interpretability (see
Section~\ref{sec:intro}). The relations are computed on all minimal
alignments, where a non-minimal representation is minimized as part of
the procedure. Interpreting the relations based on non-minimal
representations yield surprising results. When we consider the reference
$R = \texttt{GCTTT}$ with variant~$\varphi_O = \texttt{[1G>A;2C>G;3T>C]}$
($O = \texttt{AGCTT}$) and variant~$\varphi_P = \texttt{[1G>A;2C>G]}$
($P = \texttt{AGTTT}$). The naive conclusion, based on the non-minimal
representation, would be $\varphi_O$ contains $\varphi_P$. However, both
$\varphi_O$ and $\varphi_P$ are not minimal. The minimal alignments for
$\Phi(R, O) = \{\texttt{[0\_1insA;3delT]}, \texttt{[0\_1insA;4delT]},
\texttt{[0\_1insA;5delT]}\}$ and the minimal alignment for
$\Phi(R, P) = \{\texttt{[0\_1insA;2delC]}\}$ show that the actual
relation is overlap instead of containment.

\bigskip

\noindent
A variant representation (in the classical sense) that covers all
possible minimal alignments simultaneously is impossible to find in the
general case because of potential mutual exclusivity of subalignments.
A trivial solution is the full listing of the observed sequence.
This, however, offsets the benefits of a representation that is
humanly understandable and furthermore it introduces a huge amount of
redundant information for larger sequences. However, based on the
maximal influence intervals introduced in Section~\ref{sec:influence}, a
normalized \emph{supremal} variant representation can be defined. These
take the form of a deletion insertion where the deletion spans the entire
maximal influence interval and the insertion potentially contains
redundant reference information. For the SNV example in
Section~\ref{sec:general} the supremal representation
is~\texttt{2\_3delinsAT}, where first an~\texttt{A} is deleted and
inserted again. SPDI (and consequently VRS) prescribes a normalization
procedure that follows a similar approach~\cite{spdi} by extending the
variant in both directions using a rolling procedure. We note that such a
procedure, in general, does not result in all minimal alignments (nor the
extreme bounds) being contained in the representation for all variants.

Arguably, a supremal representation is not suitable in all contexts,
e.g., reporting clinical results, but within the context of storing large
quantities of variants in, for instance a database, the proposed supremal
representations are appealing as the variants can be properly ordered and
indexed on their deleted interval. Furthermore, these representations
contain all information needed to determine the relations with other
variants in the database without the need to use the reference sequence.
The drawback, however, is that potentially larger inserted sequences are
stored (\texttt{AT} in the example). In practice however, the maximal
influence intervals are tiny compared to the length of the reference
sequence.

\section{Conclusions}\label{sec:conclusion}

Looking beyond the identification of equivalent variants, we introduced a
comprehensive set of Boolean relations: equivalence, containment, overlap
and disjoint, that partitions the domain of binary variant relations.
Using these relations, additional variants of interest, i.e., variants
with a specific relation to the queried variant can be identified. We
determine these relations by taking all minimal alignments (on sequence
level) into account. The relations can be computed efficiently using a
novel algorithm that computes all minimal alignments. We have shown that
these relations occur frequently in existing datasets, notably large ones
like dbSNP. Approximately half of the variants in the \textit{CFTR}~gene
in dbSNP has at least one non-disjoint relation with another variant
within the same gene. We have shown that normalization of variant
representations is not powerful enough to answer any but the trivial
relation queries. Inspired by the alignment matrix, we introduced the
maximal influence interval of a variant. Filtering on the maximal
influence interval allows for calculating the relations for all pairs of
variants for an entire gene.

For indexing variants in a database setting, allowing querying on our
Boolean relations, we expect that the supremal representation
(Section~\ref{sec:nonminimal}) will be convenient.

In the case where phased variants (alleles) are available, directly
querying on other (combinations of) variants is possible, e.g., is a
variant contained within a given allele.
The quantification and the makeup of the overlap relation remains an
open problem. Locus specific databases can, without changing their
internal representation of variants, use our algebra to query on these
relations. Because our method is not tied to a particular representation,
it can also be applied in VCF annotation tools.

\noindent
A Python implementation is available at
\path{https://github.com/mutalyzer/algebra/tree/v0.2.0} as well as a
web interface at \path{https://mutalyzer.nl/algebra}.

\subsection{Future Work}

The current Python implementation is suitable for sequences up to a
length of that of an average gene. Preliminary work on an implementation
in a more performance oriented language indicates that our approach is
suitable for handling whole human chromosomes.
Although, from the algebra perspective, a single canonical (or
normalized) representation is insufficient, we see advantages of having
such a representation in different contexts (especially for human
interpretation). By looking at patterns within all the minimal
alignments, we can potentially construct a canonical representation that
reflects these patterns on sequence level in the variant, e.g., repeated
elements can be separated from larger variants or a sequence level
argument can be given for why close by SNVs should be (or not be)
combined. These observations could be combined in a new implementation of
a variant description extractor~\cite{vde}.

Dealing with variants in an algebraic way can possibly be extended to
higher-level calculations such as union, subtraction and
characterizing/measuring overlap. The ability to mathematically construct
larger alleles from smaller variants seems appealing in many domains.
These techniques would also enable a proper sequence level remapping of
variants onto other reference sequences, which is a recurring problem
with the publication of every new reference genome.

\printbibliography

\end{document}